\documentclass[prl,notitlepage,aps]{revtex4-2}
\usepackage{epsfig}
\usepackage{hyperref}
\usepackage{graphicx}  
\usepackage{dcolumn}   
\usepackage{bm}        
\usepackage{amsmath}
\usepackage{amsfonts}
\usepackage{bbm}
\usepackage{float}
\usepackage[caption=false]{subfig}
\usepackage{enumerate}
\usepackage{tikz}
\usepgflibrary{arrows.meta}
\usepackage{physics}
\usepackage{mathrsfs}
\usepackage[normalem]{ulem}
\usepackage{graphicx}
\usepackage[caption=false]{subfig}

\usepackage{lipsum}

\newcommand{\Ce}{\ensuremath{\text{CeRh}_{2}\text{As}_{2}}}

\begin{document}

\title{Kramers' degenerate magnetism and superconductivity}
\author{Adil Amin}
    \thanks{These authors contributed equally.}
    \author{Hao Wu}%
    \thanks{These authors contributed equally.}
\author{Tatsuya Shishidou} 
\author{Daniel F. Agterberg}

\affiliation{$^1$Department of Physics, University of Wisconsin--Milwaukee, Milwaukee, Wisconsin 53201, USA}
\date{\today}

\begin{abstract} 
Motivated by the recent discovery of odd-parity multipolar antiferromagnetic  order in \Ce, we examine the coexistence of such translation invariant Kramers' degenerate magnetic states and superconductivity. We show that the presence of such magnetic states generically suppresses superconductivity, whether it be spin-singlet or spin-triplet, unless the magnetic state drives a symmetry-required pair density wave (PDW) superconducting order. We apply our results to \Ce, where no pair density wave order appears; and to the loop current order in the cuprates, where such pair density wave superconductivity must appear together with Bogoliubov Fermi surfaces. In the former case, we explain why superconductivity is not suppressed. 
\end{abstract}

\maketitle


\noindent {\it Introduction} \ The discovery of likely field-induced odd-parity superconductivity in CeRh$_2$As$_2$ \cite{khim2021field} has generated a great deal of interest \cite{Schertenleib_2021,Ptok_2021,Moeckli_2021,Skurativska_2021,nogaki2021topological,cavanagh2022nonsymmorphic,landaeta2022field,nogaki2022even,mockli2021superconductivity,hazra2022triplet,mishra2022anisotropic,kimura2021optical,machida2022violation,semeniuk2023superconductivity,onishi2022low}. Subsequent to this, in addition to quadrupolar order \cite{landaeta2022field}, CeRh$_2$As$_2$ has been found to host odd-parity multipolar antiferromagnetic (AFM) order \cite{kibune2022observation,kitagawa2022two}. In particular, nuclear quadrupolar resonance (NQR) and nuclear magnetic resonance (NMR) have observed that this order develops at a temperature below the superconducting transition temperature ($T_c$). The symmetry of such a state 
is unusual since it independently breaks parity ($\mathcal{P}$) and time reversal 
($\mathcal{T}$) symmetries but is  invariant under the action of their 
product $\mathcal{PT}$. Since $\mathcal{PT}$ is still preserved, electronic states retain the well-known two-fold Kramers' 
degeneracy, hence we shall refer to such magnetic states  as Kramers' degenerate magnets. Such order has been referred to as odd parity multipole order~\cite{sumita2016superconductivity} and magnetic toroidal order~\cite{hayami2018classification, wu2023nematic} in different contexts.   Since neither $\mathcal{P}$ nor $\mathcal{T}$ symmetries are present, the electronic dispersion $\xi_{\boldsymbol{k}}$ does not satisfy the usual relation 
$\xi_{\boldsymbol{k}}=\xi_{-\boldsymbol{k}}$. The asymmetry between $\xi_{\boldsymbol{k}}$ and $\xi_{-\boldsymbol{k}}$ ($\xi_{\boldsymbol{k}}\neq\xi_{-\boldsymbol{k}}$) naturally opens up the question of what effect  Kramers' degenerate magnetic order has on the superconducting state. This question has been addressed  previously for multi-band systems 
\cite{sumita2016superconductivity,sumita2017multipole} and also for loop current order in cuprates ~\cite{berg2008stability,allais2012loop,wang2013quantum}. Superconductivity is usually a low energy scale phenomenon, thus it is desirable to develop a general single band theory to capture the interplay of Kramers' degenerate order and superconductivity. Here we provide such a general theoretical framework which allows us to encompass arbitrary energy dispersions and gap functions, allowing us to draw broad conclusions. 

Our theory shows that both spin-singlet and spin-triplet superconductivity are  
strongly suppressed by such an order (such states can still survive in one dimensional systems \cite{sumita2016superconductivity}). However, if the magnetic state belongs to a special symmetry class then superconductivity can still appear. In particular, if the magnetic state belongs to a vector representation of the point group, then a symmetry dictated pair density wave (PDW) superconducting  state is stabilized. Furthermore, we find this magnetic order often implies the existence 
of Fermi surfaces in the superconducting states. Such Fermi surfaces are called Bogoliubov Fermi surfaces (BFS). When the magnetic state leads to  symmetry required PDW superconductivity, we find that the PDW order reduces the size of the Bogoliubov Fermi surfaces, providing a natural microscopic mechanism for the appearance of such PDW order. We apply this framework to 
superconductivity coexisting with odd-parity multipolar AFM order in CeRh$_2$As$_2$ and coexisting with the loop current 
order in the cuprates. In our companion paper, we also apply the general framework developed to obtain a nematic BFS in the tetragonal phase of FeSe$_{1-x}$S$_x$ \cite{wu2023nematic}. Such a BFS has recently been observed in ARPES measurements for FeSe$_{1-x}$S$_x$~\cite{nagashima2022discovery}. \\

\label{sec:intro}

\noindent {\it Odd parity energy dispersion} \ As discussed above, Kramers' degenerate magnetic order gives rise to asymmetry between the electronic dispersions $\xi_{\boldsymbol{k}}$ and $\xi_{-\boldsymbol{k}}$. This asymmetry is the key ingredient in our theory and we define $\xi_{-,\boldsymbol{k}}=(\xi_{\boldsymbol{k}}-\xi_{-\boldsymbol{k}})/2$  to quantify this. The ${\boldsymbol{k}}$ dependence of $\xi_{-,\boldsymbol{k}}$ is governed by the symmetry of the Kramers' degenerate magnetic order \cite{hayami2018classification}. Since these magnetic orders are translation invariant, they correspond to odd-parity irreducible representations of the crystallographic point group. Therefore, $\xi_{-,\boldsymbol{k}}$ can be constructed by forming products of momentum components $\boldsymbol{k}_i$ that belong to the corresponding irreducible representation of the crystallographic point group. Some examples of the predicted form of $\xi_{-,\boldsymbol{k}}$ can be  found in \cite{fedchenko2022direct,vsmejkal2017electric,sumita2017multipole,sumita2016superconductivity,urru2022neutron,bhowal2021revealing,spaldin2019advances,PhysRevB.105.155157,PhysRevB.106.024405} and the recent experimental observation of this has been reported in MnAu$_2$ \cite{fedchenko2022direct}. Here we discuss what form $\xi_{-,\boldsymbol{k}}$ takes for the observed odd-parity AFM order in CeRh$_2$As$_2$ and for the loop current order in the cuprates.  

In CeRh$_2$As$_2$, with space group P4/nmm, there are two Ce atoms per unit cell. Including a single Kramers' doublet on each Ce site leads to the tight-binding Hamiltonian \cite{khim2021field},
\begin{align}
H_0= \ &\epsilon_{00, {\bm k}}\tau_0\sigma_0+\epsilon_{x0, {\bm
      k}}\tau_x\sigma_0+\epsilon_{y0, {\bm k}}\tau_y\sigma_0  
 +\epsilon_{zx, {\bm k}}\tau_z\sigma_x+\epsilon_{zy, {\bm
  k}}\tau_z\sigma_y+\epsilon_{zz, {\bm k}}\tau_z\sigma_z \notag 
  \\
  =
 \ &[t_1(\cos k_x+\cos k_y)-\mu]\tau_0\sigma_0+t_{c,1}\cos \frac{k_z}{2}\cos \frac{k_x}{2}\cos\frac{k_y}{2}\tau_x\sigma_0 
+t_{c,2}\sin \frac{k_z}{2}\cos \frac{k_x}{2}\cos\frac{k_y}{2}\tau_y\sigma_0 \notag\\
&+\alpha_R(\sin k_y \tau_z\sigma_x-\sin k_x \tau_z\sigma_y)+\lambda \sin k_z (\cos k_x-\cos k_y)\sin k_x\sin k_y\tau_z\sigma_z .
 \label{eq:H0}
\end{align}
The $\tau_i$  Pauli matrices encode the Ce site basis, and the $\sigma_i$ Pauli matrices encode the spin basis. Here $\alpha_R$ and $\lambda$ are the Rashba and Ising spin-orbit couplings. To this, we add the minimal coupling to the odd-parity AFM order. This order is of opposite sign on the two Ce sites in the unit cell and the Ce moments are oriented along the $z$ axis \cite{kibune2022observation,kitagawa2022two}. Consequently, we write this coupling as $H_c=M_z \tau_z\sigma_z$. The resulting Hamiltonian gives rise to two pseudospin (or doubly-degenerate) bands. Treating $H_c$ as a perturbation to $H_0$ yields 
\begin{equation}
\label{epsilon_odd}
\xi_{-,\boldsymbol{k}}=\pm \frac{M_z\lambda \sin k_z (\cos k_x-\cos k_y)\sin k_x\sin k_y}{\sqrt{\epsilon_{x0}^2+\epsilon_{y0}^2+\epsilon_{zx}^2+\epsilon_{zy}^2+\epsilon_{zz}^2}}, 
 \end{equation}
where the $\pm$ denotes that $\xi_{-,\boldsymbol{k}}$ has a 
different sign on the two bands. A key feature of the 
$\xi_{-,\boldsymbol{k}}$ is that even though it has the same 
form for both pseudospin partners, it is generated through the 
spin-orbit coupling term $\lambda$. Formally,  
$\xi_{-,\boldsymbol{k}}$ has a momentum structure that  belongs 
to the $A_{1u}$ representation of the $D_{4h}$ point group.

The form of $\xi_{-,\boldsymbol{k}}$ for  loop current order in 
the cuprates has been derived earlier and takes the form  
\cite{berg2008stability,varma1997non,varma1999pseudogap,simon2002detection,varma2006theory}
\begin{align}
\xi_{-,\boldsymbol{k}}= 2 J\left[\sin k_x-\sin k_y-\sin \left(k_x-k_y\right)\right].
\label{cuprate_odd}
\end{align}
where $J$ is the strength of the  order. Physically, this term originates from orbital currents that form 
closed loops within the CuO$_2$ unit cell. In this case, unlike for CeRh$_2$As$_2$ above, spin-orbit coupling does not play a role in the origin of 
$\xi_{-,\boldsymbol{k}}$. Formally, for the loop current order, the momentum structure of 
$\xi_{-,\boldsymbol{k}}$ 
belongs to the $E_{u}$ representation of the $D_{4h}$ point 
group. In contrast to the odd-parity multipole AFM
order in CeRh$_2$As$_2$, here the momentum structure of  
$\xi_{-,\boldsymbol{k}}$ belongs to a vector representation of 
the $D_{4h}$ point group. This, as we shall see later, leads to the stabilization of PDW superconductivity for the cuprates.

\noindent{\it General Theory} \  The Kramers' degeneracy that is retained with the presence of  Kramers' degenerate magnetic order allows a single band theory for superconductivity \cite{sigrist1991phenomenological,gor2001superconducting}. In the weak coupling limit, this can be formulated for arbitrary Fermi surfaces, gap functions, and  $\xi_{-,\boldsymbol{k}}$ and this further will allow this physics to be treated within the framework of quasi-classical theory, a powerful framework within which to examine superconductivity \cite{Serene:1983}. We consider the following single-band Hamiltonian 
\begin{equation}
H=\sum_{\boldsymbol{k s}} \varepsilon_{\boldsymbol{k}} c_{\boldsymbol{k s}}^{\dagger} c_{\boldsymbol{k s}}+\frac{1}{2} \sum_{\boldsymbol{k} \boldsymbol{k}^{\prime} \alpha \beta \alpha^{\prime} \beta^{\prime}} V_{\alpha \beta \alpha^{\prime} \beta^{\prime}}\left(\boldsymbol{k}, \boldsymbol{k}^{\prime}\right) c_{\boldsymbol{k}+\frac{\boldsymbol{q}}{2}, \alpha}^{\dagger} c_{-\boldsymbol{k}+\frac{\boldsymbol{q}}{2}, \beta}^{\dagger} c_{-\boldsymbol{k}^{\prime}+\frac{\boldsymbol{q}}{2}, \beta^{\prime}} c_{\boldsymbol{k}^{\prime}+\frac{\boldsymbol{q}}{2}, \alpha^{\prime}}
\end{equation}
where the operator $c_{\boldsymbol{k s}}^{\dagger}\left(c_{\boldsymbol{k s}}\right)$ creates (annihilates) electrons with momentum $\boldsymbol{k}$ and spin $s$; $s, \alpha, \beta, \alpha^{\prime}, \beta^{\prime}=\uparrow,\downarrow$; $\boldsymbol{q}$ is the center-of-mass momentum of the Cooper pairs; 
$\varepsilon_{\boldsymbol{k}} \equiv \xi_{\boldsymbol{k}}-\mu=\xi_{+, \boldsymbol{k}}+\xi_{-, \boldsymbol{k}}-\mu$, and $\xi_{+,\boldsymbol{k}}=(\xi_{\boldsymbol{k}}+\xi_{-\boldsymbol{k}})/2$ is the usual $\boldsymbol{k}$ symmetric dispersion included for single-band theories, $\mu$ is the chemical potential. The role of Kramers' degenerate magnetic order is included through the addition of  $\xi_{-,\boldsymbol{k}}$.
Here we explicitly include the Cooper center of mass momentum since later we find  that symmetry requires this must be nonzero for Kramers' degenerate magnetic order belonging to a vector representation of the point group. 

Prior to presenting the results of our analysis, we note that generally Kramers' degenerate magnetic order leads to mixing between even parity, pseudospin-singlet,  and odd parity, pseudospin-triplet, superconductivity. This can be understood from Ginzburg-Landau (GL) theory. In particular, consider an even parity order parameter $\psi$, an odd parity order parameter $\eta$, and a Kramers' degenerate magnetic order parameter $M$. Then symmetry allows the following term $iM(\psi\eta^*-\eta\psi^*)$ in the GL free energy, which ensures a singlet-triplet mixed order parameter of the form $\psi+i\eta$. However, in the single band limit we consider here, it is possible to show that this pseudospin singlet-triplet mixing vanishes if we assume that the pairing interactions are not changed by the Kramers' degenerate magnetic order. This  is justified if $\xi_{-,\boldsymbol{k}}$ has a much smaller energy scale than the electronic bandwidth. Hence we consider each pairing channel independently. For pseudospin singlet order, we take $V_{\alpha \beta \alpha^{\prime} \beta^{\prime}}\left(\boldsymbol{k}, \boldsymbol{k}^{\prime}\right)=\frac{V_s}{2}\left[f_{\boldsymbol{k}} i {\sigma}_{\mathrm{2}}\right]_{\alpha \beta}\left[f_{\boldsymbol{k}^{\prime}} i {\sigma}_{\mathrm{2}}\right]_{\alpha^{\prime} \beta^{\prime}}^{\dagger}$ while for pseudospin triplet order we take $V_{\alpha \beta \alpha^{\prime} \beta^{\prime}}\left(\boldsymbol{k}, \boldsymbol{k}^{\prime}\right)=-\frac{V_t}{2}\left[\boldsymbol{d}_{\boldsymbol{k}} \cdot {\boldsymbol{\sigma}} i {\sigma_2}\right]_{\alpha \beta}\left[\boldsymbol{d}_{\boldsymbol{k}^{\prime}} \cdot {\boldsymbol{\sigma}} i {\sigma_2}\right]_{\alpha^{\prime} \beta^{\prime}}^{\dagger}$ and assume that both $f_{\boldsymbol{k}}$ and $\boldsymbol{d}_{\boldsymbol{k}}$ can be chosen real - in practice this applies to singly degenerate superconducting irreducible representations. It is not difficult to generalize this to higher-dimensional irreducible representations.

Since we are interested in the evolution of the superconducting state both at low temperatures and near $T_c$, we compute the free energy within  mean-field theory
\begin{equation}
\Omega_{\mathrm{s}}-\Omega_{\mathrm{n}}=\left\{-2 k_B \mathrm{~T} \sum_{\boldsymbol{k}} \ln \left[\cosh \left(\frac{\beta E_{\boldsymbol{k}, \boldsymbol{q}}}{2}\right)\right]+\frac{\left|\Delta_{\boldsymbol{q}}\right|^2}{V_s}\right\} 
-\left\{-2 k_B \mathrm{~T} \sum_{\boldsymbol{k}} \ln \left[\cosh \left(\frac{\beta E_{\boldsymbol{k}, \boldsymbol{q}, 0}}{2}\right)\right]\right\}
\label{free-energy}
\end{equation}
where $E_{\boldsymbol{k}, \boldsymbol{q}}\equiv(\Delta \varepsilon)_{\boldsymbol{k}, \boldsymbol{q}}+\sqrt{\epsilon_{\boldsymbol{k}, \boldsymbol{q}}^2+\left|\Delta_{\boldsymbol{q}} f_{\boldsymbol{k}}\right|^2}$ with $(\Delta \varepsilon)_{\boldsymbol{k}, \boldsymbol{q}} \equiv \frac{1}{2}\left(\varepsilon_{\boldsymbol{k}+\frac{\boldsymbol{q}}{2}}-\varepsilon_{-\boldsymbol{k}+\frac{\boldsymbol{q}}{2}}\right)$ and $\epsilon_{\boldsymbol{k}, \boldsymbol{q}}\equiv \frac{1}{2}\left(\varepsilon_{\boldsymbol{k}+\frac{\boldsymbol{q}}{2}}+\varepsilon_{-\boldsymbol{k}+\frac{\boldsymbol{q}}{2}}\right)$, $E_{\boldsymbol{k}, \boldsymbol{q}, 0}\equiv(\Delta \varepsilon)_{\boldsymbol{k}, \boldsymbol{q}}+\sqrt{\epsilon_{\boldsymbol{k}, \boldsymbol{q}}^2}$, $V_s$ is the pairing interaction. Eq.~\ref{free-energy} applies to pseudospin-singlet order, for the pseudospin-triplet order, $\left|f_{\boldsymbol{k}}\right|^2$ is replaced by $\left|\boldsymbol{d}_{\boldsymbol{k}}\right|^2$.
Minimizing this free energy with respect to 
$\left|\Delta_{\boldsymbol{q}}\right|$ and $\boldsymbol{q}$ leads to the gap equation and the condition on $\boldsymbol{q}$ that ensures a vanishing  supercurrent
\begin{equation}
\ln \left(\frac{T}{T_{c 0}}\right)=\left\langle 2 \pi k_B T\left|f_{\boldsymbol{k}}\right|^2 R e \sum_{n=0}^{\infty}\left(\frac{1}{\delta}-\frac{1}{\hbar \omega_n}\right)\right\rangle_k
\end{equation}
\begin{equation}
\left\langle\frac{\partial(\Delta \varepsilon)_{\boldsymbol{k}, \boldsymbol{q}}}{\partial \boldsymbol{q}}\left|f_{\boldsymbol{k}}\right|^2 \operatorname{Im} \sum_{n=0}^{\infty} \frac{1}{\delta\left(\delta+\hbar \omega_n+i(\Delta \varepsilon)_{\boldsymbol{k}, \boldsymbol{q}}\right)}\right\rangle_k=0
\end{equation}
where $\delta \equiv \sqrt{\left(\hbar \omega_n+i(\Delta \varepsilon)_{\boldsymbol{k}, \boldsymbol{q}}\right)^2+\left|\Delta_{\boldsymbol{q}}\right|^2\left|f_k\right|^2}$, and $\omega_n$ is the Matsubara frequency which satisfies $\hbar \omega_n=(2 n+1) \pi k_B T, n \in \mathbb{Z}$. Throughout this work, we adopt the convention that the average over the Fermi surface $\left\langle\left|f_k\right|^2\right\rangle_k=1$. This average over the Fermi surface is weighted by the momentum-dependent density of states. By numerically solving this set of equations, we can find the gap and the optimal $\boldsymbol{q}$ as functions of $T$.
As shown in the Appendix, the Green's function and the anomalous Green's function are
\begin{equation}
\hat{G}=-\frac{\hbar\left(\varepsilon_{-\boldsymbol{k}+\frac{\boldsymbol{q}}{2}}+i \hbar \omega_n\right) \sigma_0}{\left(\varepsilon_{-\boldsymbol{k}+\frac{\boldsymbol{q}}{2}}+i \hbar \omega_n\right)\left(\varepsilon_{\boldsymbol{k}+\frac{\boldsymbol{q}}{2}}-i \hbar \omega_n\right)+|\psi(\boldsymbol{k}, \boldsymbol{q})|^2}
\end{equation}
\begin{equation}
\hat{F}=\frac{\hbar \hat{\Delta}}{\left(\varepsilon_{-\boldsymbol{k}+\frac{\boldsymbol{q}}{2}}-i \hbar \omega_n\right)\left(\varepsilon_{\boldsymbol{k}+\frac{\boldsymbol{q}}{2}}+i \hbar \omega_n\right)+|\psi(\boldsymbol{k}, \boldsymbol{q})|^2}
\end{equation}
where $\sigma_0$ is the identity matrix and $\hat{\Delta}$ is the gap matrix. As can be also seen in the Appendix, for the pseudospin-triplet $|\psi(\boldsymbol{k}, \boldsymbol{q})|^2$ is replaced by $|\boldsymbol{d}(\boldsymbol{k}, \boldsymbol{q})|^2$

To gain an understanding of the role of the Kramers' degenerate magnetic order, we initially consider $T$ near  $T_c$. The equation determining the critical temperature $T_c^q$ is
\begin{equation}
\ln \left(\frac{T_c^q}{T_{c 0}}\right)  
=\left\langle\left|f_k\right|^2\left\{\psi\left(\frac{1}{2}\right)-R e\left[\psi\left(\frac{1}{2}+\frac{i (\Delta \varepsilon)_{\boldsymbol{k}, \boldsymbol{q}}
}{2\pi k_B T_c^q}\right)\right]\right\}\right\rangle_k
\end{equation}
where $T_{c 0}$ is the critical temperature for $(\Delta \varepsilon)_{\boldsymbol{k}, \boldsymbol{q}}=0$, the digamma function $\psi(z)=-\gamma+\sum_{n=0}^{\infty}\left(\frac{1}{n+1}-\frac{1}{n+z}\right)$, and $\gamma$ is the Euler-Mascheroni constant. Assuming $|\boldsymbol{q}| \ll|\boldsymbol{k}|$, we have $(\Delta \varepsilon)_{\boldsymbol{k}, \boldsymbol{q}}=\xi_{-,\boldsymbol{k}}+\frac{\boldsymbol{q}}{2} \cdot \boldsymbol{\nabla} \xi_{+,\boldsymbol{k}}$. The expression $\psi\left(\frac{1}{2}\right)-R e\left[\psi\left(\frac{1}{2}+\frac{i (\Delta \varepsilon)_{\boldsymbol{k}, \boldsymbol{q}}}{2 \pi k_B T_c^q}\right)\right]$ is intrinsically negative, which yields $T_c^q<T_{c 0}$. From the monotonic decreasing dependence of the Digamma function, we find that $T_c^q$ is decreasing with increasing $|(\Delta \varepsilon)_{\boldsymbol{k}, \boldsymbol{q}}|$. Eventually, $T_c^q$ will become zero, i.e., superconductivity is destroyed. Using the asymptotic expansion of the Digamma function, the critical value for $|(\Delta \varepsilon)_{\boldsymbol{k}, \boldsymbol{q}}|$ corresponding to $T_c=0$ is given by
\begin{equation}
\left\langle\left|f_{\boldsymbol{k}}\right|^2 \ln \left(\frac{|(\Delta \varepsilon)_{\boldsymbol{k}, \boldsymbol{q}}
|}{2 \pi k_B T_{c 0}}\right)\right\rangle_k=\psi\left(\frac{1}{2}\right)
\end{equation}
We conclude that once the magnitude of $(\Delta \varepsilon)_{\boldsymbol{k}, \boldsymbol{q}}$ becomes on the order of $k_B T_{c 0}$  superconductivity is destroyed. We note that the usual Pauli limiting field ,$H_p$,  for an $s$-wave superconductor is given by $\mu_BH_P=\Delta_0/\sqrt{2}$ hence, throughout this paper to we choose to express $(\Delta \varepsilon)_{\boldsymbol{k}, \boldsymbol{q}}$ in units of $\mu_BH_P$ to provide some context.  Importantly, since $(\Delta \varepsilon)_{\boldsymbol{k}, \boldsymbol{q}}$ depends not only on $\xi_{-,\boldsymbol{k}}$, but also on  
  $\boldsymbol{q}$, it is possible that $(\Delta \varepsilon)_{\boldsymbol{k}, \boldsymbol{q}}$ can become small even if $\xi_{-,\boldsymbol{k}}$ is larger than the superconducting gap. To illustrate this, it is useful to consider an example with $\xi_{+,\boldsymbol{k}}=\frac{\hbar^2 \boldsymbol{k}^2}{2 m}$ and $\xi_{-,\boldsymbol{k}}=\alpha k_x$. Then, $(\Delta \varepsilon)_{\boldsymbol{k}, \boldsymbol{q}}=\alpha k_x+\frac{\hbar^2}{2 m}\left(q_x k_x+q_y k_y\right)$. The choice $\boldsymbol{q}=\left(q_x, q_y\right)=\left(-\frac{2 m \alpha}{\hbar^2}, 0\right)$ will cancel the $\xi_{-,\boldsymbol{k}}$ term making $(\Delta \varepsilon)_{\boldsymbol{k}, \boldsymbol{q}}=0$. Therefore, the critical temperature remains unchanged from $T_{c 0}$ in spite of a $\xi_{-,\boldsymbol{k}}$ that can be much larger than the gap.

 To understand in more detail how a PDW state with a non-zero $\boldsymbol{q}$ can resurrect a suppressed $T_c$ due to $\xi_{-,\boldsymbol{k}}$, we carry out a small $(\Delta \varepsilon)_{\boldsymbol{k}, \boldsymbol{q}}$ expansion. The critical temperature is then determined by
\begin{equation}
\ln \left(\frac{T_c^q}{T_{c 0}}\right)=\frac{7 \zeta(3)}{\left(2 \pi k_B
T_c^q\right)^2}    \notag \\
\times \left\langle\left|f_{\boldsymbol{k}}\right|^2\left[-\xi_{-,\boldsymbol{k}}
^2-\frac{\hbar^2(\boldsymbol{q} \cdot \boldsymbol{v})^2}{4}-\hbar \xi_{-,\boldsymbol{k}}
\boldsymbol{q} \cdot \boldsymbol{v}\right]\right\rangle_k
\end{equation}
where $\boldsymbol{v}(\boldsymbol{k}) \equiv \frac{1}{\hbar} \nabla \xi_{+,\boldsymbol{k}}$ is the velocity. Only the third term ($-\hbar \xi_{-,\boldsymbol{k}}\boldsymbol{q} \cdot \boldsymbol{v}$) can be positive and hence increase $T_c^q$. Thus, the existence of a PDW state depends on a nonzero average $\left\langle\left|f_{\boldsymbol{k}}\right|^2\left[-\hbar \xi_{-,\boldsymbol{k}}\boldsymbol{q} \cdot \boldsymbol{v}\right]\right\rangle_k$. Since the components of the velocity $\boldsymbol{v}$ belong to a vector representation of the point group, the only way to get this average nonzero is to require that the momentum structure of $\xi_{-,\boldsymbol{k}}$ also belongs to a vector representation. This implies that the Kramers' degenerate magnetic order belongs to a vector representation of the point group. Indeed, when the Kramers' degenerate magnetic order belongs to a vector representation, symmetry implies that the Ginzburg-Landau free energy contains a non-vanishing Lifshitz invariant of the form $\sum_j \gamma_j\left[\psi\left(D_j \psi\right)^*+\psi^*\left(D_j \psi\right)\right]$ with $D_j \equiv-i \nabla_j-2 e A_j$, which guarantees the appearance of a PDW state with non-zero $\boldsymbol{q}$. This symmetry dictated PDW order typically takes the form $\psi=\psi_0 e^{i 
\boldsymbol{q} \cdot \boldsymbol{r}}$ with a single-plane wave, unlike the multiple plane-wave solutions often associated with PDW order \cite{Agterberg:2020}.  The existence of Lifshitz invariants, and the concomitant finite momentum PDW pairing, is important for the superconducting diode effect \cite{yuan2022supercurrent, he2022phenomenological, daido2022intrinsic, pal2022josephson}. This suggests that superconductivity coexisting with the Kramers' degenerate magnetic order provides a route towards creating the superconducting diode effect.

In addition to the generating PDW states, Kramers' degenerate magenetic order can also lead to novel low energy excitation spectra. In particular, by finding the poles of the Green's function $G$, we can obtain the quasiparticle dispersion
\begin{equation}
E=(\Delta \varepsilon)_{\boldsymbol{k}, \boldsymbol{q}}
 \pm \sqrt{\epsilon_{\boldsymbol{k}, \boldsymbol{q}}^2+\left|\Delta_{\boldsymbol{q}} f_{\boldsymbol{k}}\right|^2}
\end{equation}
In principle, because of the appearance of $(\Delta \varepsilon)_{\boldsymbol{k}, \boldsymbol{q}}$ in this expression, Bogoliubov Fermi 
surfaces (BFS) can exist. These BFS are given by 
$E=0$ and appear for $\boldsymbol{k}$ where $(\Delta \varepsilon)_{\boldsymbol{k}, \boldsymbol{q}}$ is larger than $|\Delta_{\boldsymbol{q}} f_{\boldsymbol{k}}|$. Indeed, these are guaranteed to appear when there are gap nodes present i.e. $f_{\boldsymbol{k}}=0$ (or $\left|\boldsymbol{d}_{\boldsymbol{k}}\right|=0$)
provided that $(\Delta \varepsilon)_{\boldsymbol{k}, \boldsymbol{q}}$ does not also vanish on these nodes. When this 
occurs, $(\Delta \varepsilon)_{\boldsymbol{k}, \boldsymbol{q}}$ inflates the nodes of the 
original gap functions into Bogoliubov Fermi surfaces
\cite{agterberg2017bogoliubov,brydon2018bogoliubov,link2020bogoliubov,timm2017inflated}.  In our companion paper \cite{wu2023nematic} we carry out a detailed analysis of the existence of Bogoliubov Fermi surfaces in FeSe$_{1-x}$S$_x$. \\

\begin{figure} 
\centering
\includegraphics[height=7cm]{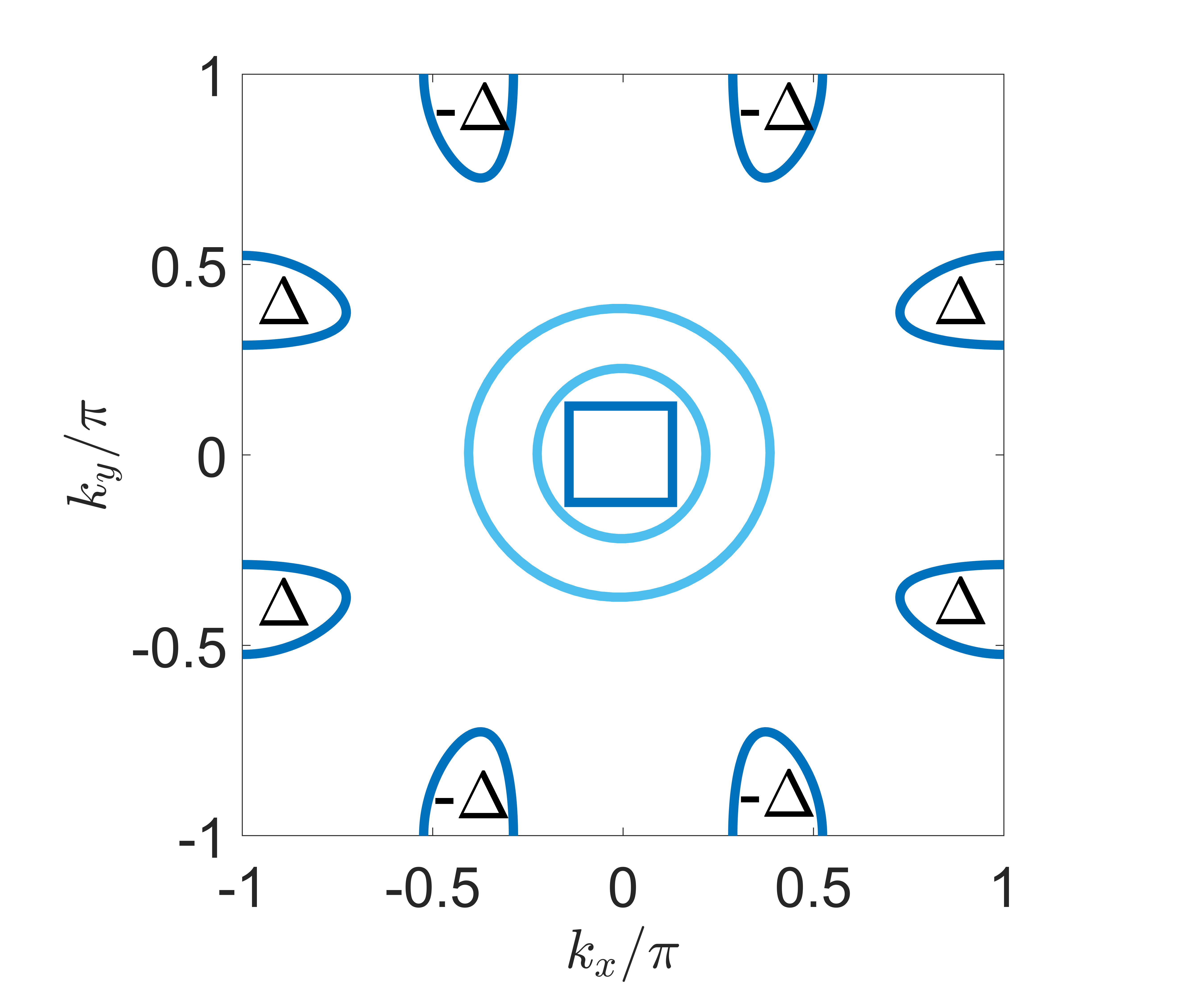}
\caption{\label{fig:Normal_FS} Sketch of the normal state Fermi surface of CeRh$_{2}$As$_{2}$ for $k_z=0$ found from renormalized DFT calculations. The majority of the density of states sit on the beans. The $d_{x^2-y^2}$ pairing symmetry considered here is denoted by the $\Delta$ values as shown.}
\end{figure}

\noindent {\it Application to CeRh$_2$As$_2$}  \ In addition to  Eq.~\ref{epsilon_odd} for $\xi_{-,\boldsymbol{k}}$, we require a description of the normal state to apply our results to CeRh$_2$As$_2$. DFT results \cite{cavanagh2022nonsymmorphic,hafner2022possible}  yield a Fermi surface as depicted in Fig.~\ref{fig:Normal_FS}. They further reveal that the majority of the density  of states (DOS) ($\sim 80\%$) is concentrated on the Fermi surface structures near the X--M line, which we refer to as beans. Here we include only those beans and, for simplicity, ignore the $c$-axis dependence of  $\xi_{+_,\boldsymbol{k}}$ (this assumption does not qualitatively change our results). Furthermore, we assume that a $d_{x^2-y^2}$ gap structure as depicted in Fig.~\ref{fig:Normal_FS} appears when no magnetic field is applied. This assumption also does not qualitatively change the results unless the gap function has nodes on the beans. We briefly comment on this possibility later. 
To describe theses beans, we develop a power series expansion in powers of  $\delta k_x,\delta k_y$ centered about the minima of the 
band along the X--M line. Explicitly considering a bean centered at $(k_x,k_y)=\left(\frac{\pi}{a}, \frac{0.4 \pi}{a}\right)$, we have the dispersion 
\begin{equation}
\xi_{+,\boldsymbol{k}}
=\alpha\left(\delta k_{x}\right)^2+\beta\left(\delta k_{y}\right)^2 
+\gamma_1\left(\delta k_{x}\right)^2\left(\delta k_{y}\right)+\gamma_2\left(\delta k_{y}\right)^3.
\end{equation} Mirror symmetry prevents terms odd in $\delta k_x$ appearing in this expression. We also expand $\xi_{-,\boldsymbol{k}}$ in Eq.~ \ref{epsilon_odd} around the centers of the beans. For the pocket centered at $\left(\frac{\pi}{a}, \frac{0.4 \pi}{a}\right), \xi_{-,\boldsymbol{k}}=$ $\Tilde{\lambda} \delta k_{x} a \sin k_z $.

\begin{figure}
\centering
\includegraphics[height=7cm]{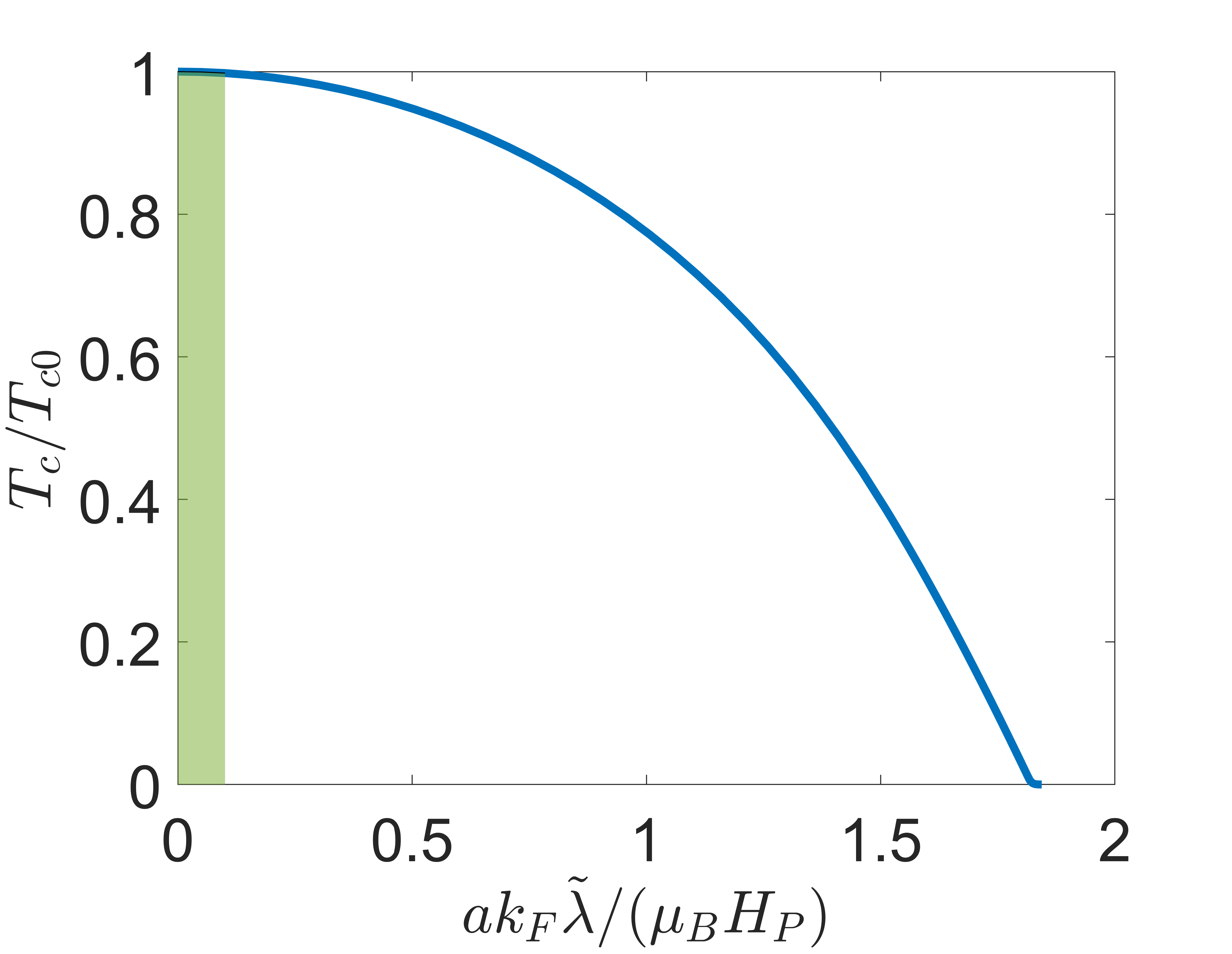}
\caption{\label{fig:Tc_lambda} Transition temperature as a function of the antisymmetric coupling parameter $\lambda$ in CeRh$_{2}$As$_{2}$. The parameter $a$ is the lattice constant,  $k_F\equiv\sqrt{\frac{\mu}{\sqrt{\alpha \beta}}}$ is the natural unit for $\delta k_x$ and $\delta k_y$. We choose the values $\alpha=\frac{1}{(1.25)^2}$, $\beta=\frac{1}{(0.55)^2}$, $\mu=0.7$, and $\gamma_1=0.75$, $\gamma_2=-0.9$. The Pauli limiting field energy scale $\mu_BH_P=\Delta_0/\sqrt{2}$, where $\Delta_0 = 1.764 k_BT_{c0}$.}
\end{figure}

The numerical solution of the linearized gap equation is shown in Fig.~\ref{fig:Tc_lambda}. Because the Kramers' degenerate magnetic order does not belong to a vector representation, a  PDW  state is not required by symmetry in this case. Thus when $\xi_{-,\boldsymbol{k}}$  becomes sufficiently large, superconductivity is suppressed as seen in  Fig. \ref{fig:Tc_lambda}. However,  this is not observed in experiment 
\cite{kibune2022observation,kitagawa2022two}, which raises the question, how does 
superconductivity still survive. Here we suggest that  the explanation 
for this persistence hinges on the value of $\Tilde{\lambda}$ defined above. $\Tilde{\lambda}$ on the beans is determined by the ratio of the Ising coupling ($\lambda$) to the Rashba spin-orbit coupling ($\alpha_R$) which occur in Eq. \ref{eq:H0}. In particular, using effective $g$ factors from  DFT based renormalized band structure calculations we constrain $\Tilde{\lambda} < 0.2 M_z $  \cite{Tatsuya}. The critical temperature of $M_z$ is of the order of  the critical temperature of superconductivity, hence $M_z \approx \mu_B H_P$ thus $\Tilde{\lambda} < 0.2 \mu_B H_P $. We further estimate $a\text{k}_F \approx \frac{1}{2}$. Hence  $\frac{a\text{k}_F\Tilde{\lambda}}{\mu_B H_P} < 0.1 $. This is shown as the shaded region in 
Fig.~ \ref{fig:Tc_lambda}. Thus superconductivity still persists despite the presence of  Kramers' degenerate magnetic order.  

 While the pairing state we consider above has $d_{x^2-y^2}$ symmetry, it is nodeless on the Fermi surface beans which do not intersect the $d_{x^2-y^2}$  nodes. However, the  $d_{x^2-y^2}$ gap will have nodes along on the $\Gamma$-centered  Fermi surfaces, possibly allowing for Bogoliubov Fermi surfaces to appear in the superconducting state. However, these nodes will not be inflated to  Bogoliubov Fermi surfaces  since $(\Delta \varepsilon)_{\boldsymbol{k}}$ vanishes along the diagonal the nodes. We  can further consider other gap functions with nodes to see if Bogoliubov Fermi surfaces are expected. For example, nodes can appear on the beans for a $d_{xy}$ like gap. The beans will then have nodes where they intersect the Brillouin Zone face. However, these nodes will also not be inflated to form Bogoliubov Fermi surfaces because again $(\Delta \varepsilon)_{\boldsymbol{k}}$ vanishes on this face. Consequently, we do not expect Bogoliubov Fermi surfaces to appear as a consequence of the Kramer's degenerate magnetic order in CeRh$_2$As$_2$ for any pairing symmetry.
 \\

\begin{figure}
\centering
\includegraphics[height=7cm]{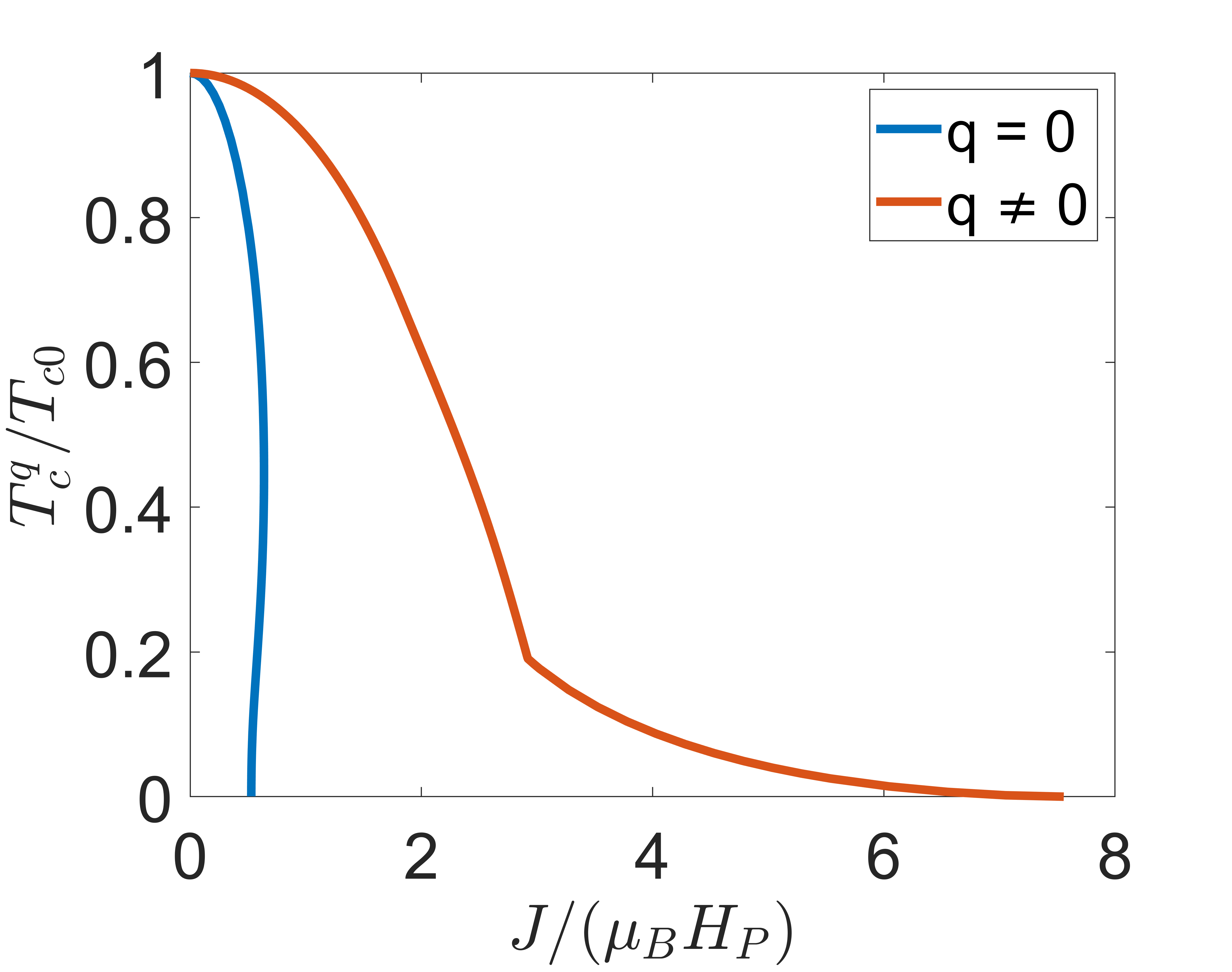}
\caption{\label{fig:Tc_J} Transition temperature as a function of the loop current order parameter for both $\boldsymbol{q}=0$ and finite $\boldsymbol{q}$ theories. The Pauli limiting field energy scale $\mu_BH_P=\Delta_0/\sqrt{2}$, where $\Delta_0 = 1.764 k_BT_{c0}$. Using hopping parameter $t$ as the energy scale, we choose $t^{\prime}=-0.25$, $\mu=-0.9$. We can observe a nonmonotonic behavior in $\boldsymbol{q}=0$ plot. This kind of behavior typically suggests the existence of a PDW state.}
\end{figure}

\begin{figure}
\centering
\includegraphics[height=10cm]{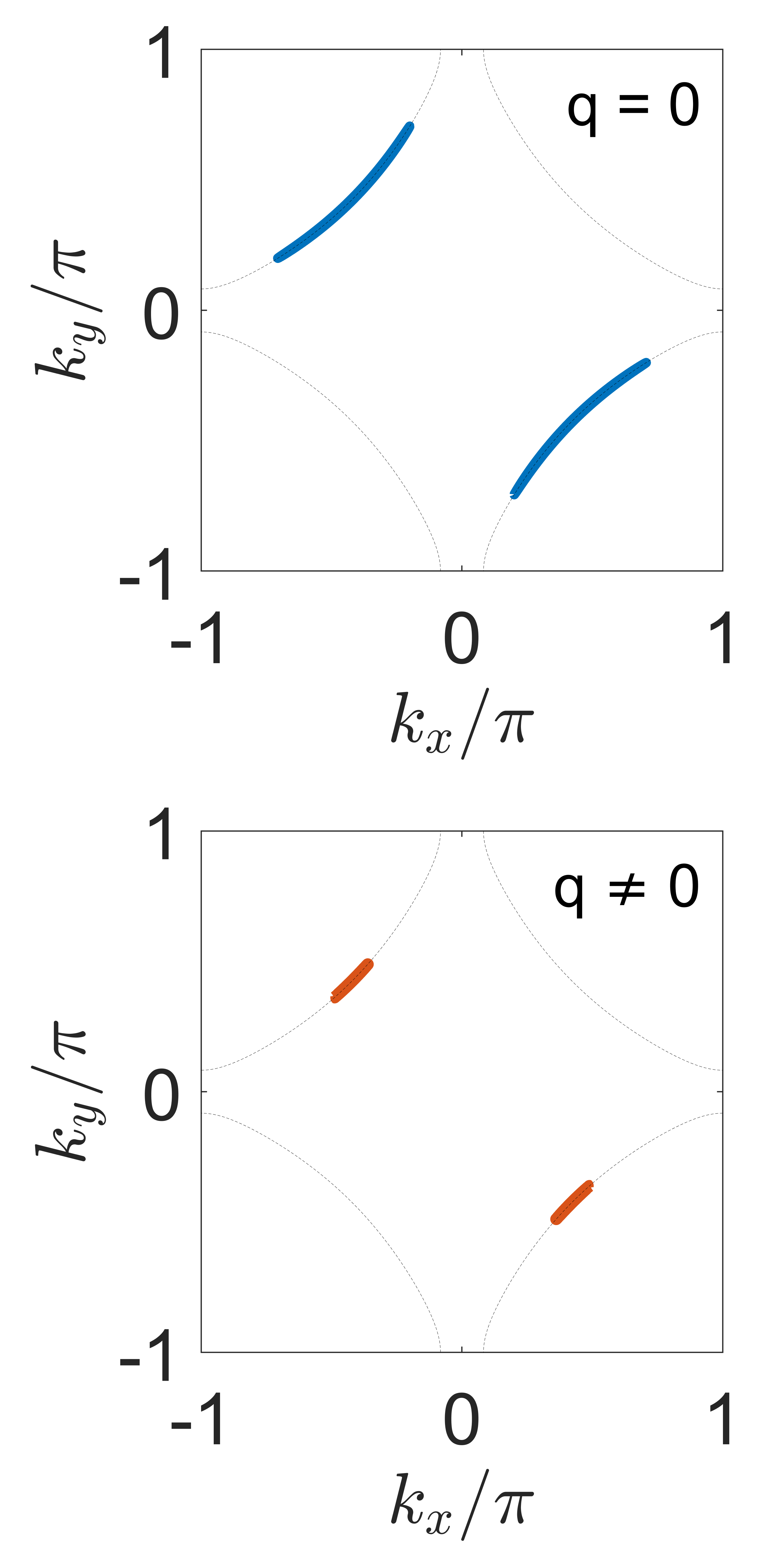}
\caption{\label{fig:BFS} Bogoliubov Fermi surfaces. The upper plot is for $q=0$, the lower plot is for $q\ne0$.}
\end{figure}

\noindent {\it Application to loop current order in cuprates} \ The normal state is given by the dispersion 
$\xi_{+,\boldsymbol{k}}=-2 t\left(\cos k_x+\cos k_y\right)-4 
t^{\prime} \cos \left(k_x\right) \cos \left(k_y\right)$, 
where $t$ and $t'$ are 
the nearest and next nearest neighbor hoppings on the square 2D lattice \cite{berg2008stability}. Eq. \ref{cuprate_odd} above gives $\xi_{-,\boldsymbol{k}}$. The gap function is taken to be $\Delta_{\boldsymbol{k}, \boldsymbol{q}}=\Delta_{\boldsymbol{q}} f_{\boldsymbol{k}}=\Delta_{\boldsymbol{q}} C\left(\cos k_x-\cos k_y\right)$ with normalization factor $C$ (determined by $\left\langle\left|f_k\right|^2\right\rangle_k=1$). Since the Kramers' degenerate magnetic order belongs to a vector representation, our general results imply that a PDW state is stabilized by the loop current order. In  Fig.~\ref{fig:Tc_J}, we compare the  calculated values for $T_c$ with and  without the PDW order. 
Remarkably, the maximum $J$ value for the PDW state is about 15 times larger than the  $\boldsymbol{q}=0$ state.
In addition, since the $d$-wave gap has nodes, the loop current order will give rise to BFSs. This has been examined previously when there is no PDW order \cite{berg2008stability,kivelson2012fermi,wang2013quantum}.  As shown in Fig.~\ref{fig:BFS}, we find that the BFS  are strongly shrunk due to the PDW order. The gain in condensation energy due to the shrinking of the BFS provides an energetic mechanism to stabilize the PDW state. 
 \\ 

\noindent {\it Conclusions} 
We considered the interplay of $\mathcal{PT}$ translation invariant Kramers' degenerate magnetic order and superconductivity.  The absence of $\mathcal{P}$ and $\mathcal{T}$ symmetry leads to an asymmetry in the electronic dispersion relation ($\xi_{\textbf{k}}\neq \xi_{-\textbf{k}}$). We develop a general single-band framework to capture the effect of the Kramers' degenerate magnetic order on superconductivity. We find that when the magnetic states belong to a vector representation of the point group, pair density wave states are stabilized. In the absence of such symmetries, superconductivity is generically suppressed. We apply our framework to \Ce, where we explain why superconductivity still persists in the presence of Kramers' degenerate magnetic order, as well as the cuprates where pair density wave superconductivity appears accompanied with Bogoliubov Fermi surfaces.\\ 

\noindent {\it Acknowledgments} A.A., H.W., and D.F.A. acknowledge support through DE-SC0021971. T.S. acknowleges support by the US Department of Energy, Office of Basic Energy Sciences, Division of Materials Sciences and Engineering under Award DE-SC0017632. We would like to thank Erez Berg, Elena Hassinger,  Seunghyun Khim, Steve Kivelson, Christian Parsons, and Michael Weinert, for useful discussions.

\bibliography{biblio}

\appendix

\section*{Appendices}

\setcounter{equation}{0}

\setcounter{subsection}{0}

\renewcommand{\theequation}{A.\arabic{equation}}

\renewcommand{\thesubsection}{A.\arabic{subsection}}

\section{Appendix A: Derivation of the Green's functions}
Starting from the single-band Hamiltonian
\begin{equation}
H=\sum_{\boldsymbol{k s}} \varepsilon_{\boldsymbol{k}} c_{\boldsymbol{k s}}^{\dagger} c_{\boldsymbol{k s}}+\frac{1}{2} \sum_{\boldsymbol{k} \boldsymbol{k}^{\prime} \alpha \beta \alpha^{\prime} \beta^{\prime}} V_{\alpha \beta \alpha^{\prime} \beta^{\prime}}\left(\boldsymbol{k}, \boldsymbol{k}^{\prime}\right) c_{\boldsymbol{k}+\frac{\boldsymbol{q}}{2}, \alpha}^{\dagger} c_{-\boldsymbol{k}+\frac{\boldsymbol{q}}{2}, \beta}^{\dagger} c_{-\boldsymbol{k}^{\prime}+\frac{\boldsymbol{q}}{2}, \beta^{\prime}} c_{\boldsymbol{k}^{\prime}+\frac{\boldsymbol{q}}{2}, \alpha^{\prime}}
\end{equation}
within mean-filed approximation, we get the mean-field Hamiltonian
\begin{equation}
H=\sum_{\boldsymbol{k s}} \varepsilon_{\boldsymbol{k}} c_{\boldsymbol{k s}}^{\dagger} c_{\boldsymbol{k s}}-\frac{1}{2} \sum_{\boldsymbol{k} \alpha \beta} \Delta_{\alpha \beta}(\boldsymbol{k}, \boldsymbol{q}) c_{\boldsymbol{k}+\frac{\boldsymbol{q}}{2}, \alpha}^{\dagger} c_{-\boldsymbol{k}+\frac{\boldsymbol{q}}{2}, \beta}^{\dagger}-\frac{1}{2} \sum_{\boldsymbol{k} \alpha \beta} \Delta_{\alpha \beta}^*(\boldsymbol{k}, \boldsymbol{q}) c_{-\boldsymbol{k}+\frac{\boldsymbol{q}}{2}, \beta} c_{\boldsymbol{k}+\frac{\boldsymbol{q}}{2}, \alpha}
\end{equation}
where we have defined the gap function as
\begin{equation}
\Delta_{\alpha \beta}(\boldsymbol{k}, \boldsymbol{q}) \equiv-\sum_{\boldsymbol{k}^{\prime} \alpha^{\prime} \beta^{\prime}} V_{\alpha \beta \alpha^{\prime} \beta^{\prime}}\left(\boldsymbol{k}, \boldsymbol{k}^{\prime}\right)\left\langle c_{-\boldsymbol{k}^{\prime}+\frac{\boldsymbol{q}}{2}, \beta^{\prime}} c_{\boldsymbol{k}^{\prime}+\frac{\boldsymbol{q}}{2}, \alpha^{\prime}}\right\rangle
\end{equation}
and then 
\begin{equation} \label{eq:gap star}
\Delta_{\alpha \beta}^*(\boldsymbol{k}, \boldsymbol{q})=-\sum_{\boldsymbol{k}^{\prime} \alpha^{\prime} \beta^{\prime}}\left[V_{\alpha \beta \alpha^{\prime} \beta^{\prime}}\left(\boldsymbol{k}, \boldsymbol{k}^{\prime}\right)\right]^*\left\langle c_{-\boldsymbol{k}^{\prime}+\frac{\boldsymbol{q}}{2}, \beta^{\prime}} c_{\boldsymbol{k}^{\prime}+\frac{\boldsymbol{q}}{2}, \alpha^{\prime}}\right\rangle^*=-\sum_{\boldsymbol{k}^{\prime} \alpha^{\prime} \beta^{\prime}} V_{\alpha^{\prime} \beta^{\prime} \alpha \beta}\left(\boldsymbol{k}^{\prime}, \boldsymbol{k}\right)\left\langle c_{\boldsymbol{k}^{\prime}+\frac{\boldsymbol{q}}{2}, \alpha^{\prime}}^{\dagger} c_{-\boldsymbol{k}^{\prime}+\frac{\boldsymbol{q}}{2}, \beta^{\prime}}^{\dagger}\right\rangle
\end{equation}
where $\left\langle ... \right\rangle$ stands for a grand canonical ensemble average.
Introduce the finite-temperature Green's function
\begin{equation}
G_{\alpha \beta}(\boldsymbol{k}, \boldsymbol{q}, \tau) \equiv-\left\langle T_\tau c_{\boldsymbol{k}+\frac{\boldsymbol{q}}{2}, \alpha}(\tau) c_{\boldsymbol{k}+\frac{\boldsymbol{q}}{2}, \beta}^{\dagger}(0)\right\rangle
\end{equation}
The operator $A(\tau)$ denotes $e^{H \tau / \hbar} A e^{-H \tau / \hbar}$, $A(0)$ is for $\tau=0$, where $H$ is the Hamiltonian, $\tau$ is the `imaginary time' which is treated as a real quantity. $T_\tau$ is the time ordering operator with respect to the imaginary time. Using the Heaviside step function 
\begin{equation}
\theta(x)= \begin{cases}0 & x<0 \\ 1 & x>0\end{cases}
\end{equation}
we can express the Green's function as 
\begin{equation}
G_{\alpha \beta}(\boldsymbol{k}, \boldsymbol{q}, \tau)=-\theta(\tau)\left\langle c_{\boldsymbol{k}+\frac{\boldsymbol{q}}{2}, \alpha}(\tau) c_{\boldsymbol{k}+\frac{\boldsymbol{q}}{2}, \beta}^{\dagger}(0)\right\rangle+\theta(-\tau)\left\langle c_{\boldsymbol{k}+\frac{\boldsymbol{q}}{2}, \beta}^{\dagger}(0) c_{\boldsymbol{k}+\frac{\boldsymbol{q}}{2}, \alpha}(\tau)\right\rangle
\end{equation}
Using $\frac{\partial}{\partial \tau} G_{\alpha \beta}(\boldsymbol{k}, \boldsymbol{q}, \tau)$, we get the equation of motion for the Green's function
\begin{equation}
\left(\frac{\partial}{\partial \tau}+\frac{\varepsilon_{\boldsymbol{k}+\frac{\boldsymbol{q}}{2}}}{\hbar}\right) G_{\alpha \beta}(\boldsymbol{k}, \boldsymbol{q}, \tau)+\frac{1}{\hbar} \sum_{\alpha^{\prime \prime}} \Delta_{\alpha \alpha^{\prime \prime}}(\boldsymbol{k}, \boldsymbol{q})\left\langle T_\tau c_{-\boldsymbol{k}+\frac{\boldsymbol{q}}{2}, \alpha^{\prime \prime}}^{\dagger}(\tau) c_{\boldsymbol{k}+\frac{\mathfrak{\boldsymbol{q}}}{2}, \beta}^{\dagger}(0)\right\rangle=-\delta(\tau) \delta_{\alpha \beta}
\end{equation}
We can define the anomalous Green's function
\begin{equation}
F_{\alpha \beta}^{\dagger}(\boldsymbol{k}, \boldsymbol{q}, \tau) \equiv-\left\langle T_\tau c_{-\boldsymbol{k}+\frac{\boldsymbol{q}}{2}, \alpha}^{\dagger}(\tau) c_{\boldsymbol{k}+\frac{\boldsymbol{q}}{2}, \beta}^{\dagger}(0)\right\rangle
\end{equation}
Similarly, using $\frac{\partial}{\partial \tau} F_{\alpha \beta}^{\dagger}(\boldsymbol{k}, \boldsymbol{q}, \tau)$, we get the equation of motion for the anomalous Green's function
\begin{equation}
\left(\frac{\partial}{\partial \tau}-\frac{\varepsilon_{-\boldsymbol{k}+\frac{\boldsymbol{q}}{2}}}{\hbar}\right) F_{\alpha \beta}^{\dagger}(\boldsymbol{k}, \boldsymbol{q}, \tau)-\frac{1}{\hbar} \sum_{\alpha^{\prime \prime}} \Delta_{\alpha^{\prime \prime} \alpha}^*(\boldsymbol{k}, \boldsymbol{q}) G_{\alpha^{\prime \prime} \beta}(\boldsymbol{k}, \boldsymbol{q}, \tau)=0
\end{equation}
Transforming the Green's function and the anomalous Green's function from the imaginary-time-momentum space to the frequency-momentum space, we get the Gor'kov equations
\begin{equation}
\left(-i \omega_n+\frac{\varepsilon_{\boldsymbol{k}+\frac{\boldsymbol{q}}{2}}}{\hbar}\right) G_{\alpha \beta}\left(\boldsymbol{k}, \boldsymbol{q}, \omega_n\right)-\frac{1}{\hbar} \sum_{\alpha^{\prime \prime}} \Delta_{\alpha \alpha^{\prime \prime}}(\boldsymbol{k}, \boldsymbol{q}) F_{\alpha^{\prime \prime} \beta}^{\dagger}\left(\boldsymbol{k}, \boldsymbol{q}, \omega_n\right)=-\delta_{\alpha \beta}
\end{equation}
\begin{equation}
\left(-i \omega_n-\frac{\varepsilon_{-\boldsymbol{k}+\frac{\boldsymbol{q}}{2}}}{\hbar}\right) F_{\alpha \beta}^{\dagger}\left(\boldsymbol{k}, \boldsymbol{q}, \omega_n\right)-\frac{1}{\hbar} \sum_{\alpha^{\prime \prime}} \Delta_{\alpha^{\prime \prime} \alpha}^*(\boldsymbol{k}, \boldsymbol{q}) G_{\alpha^{\prime \prime} \beta}\left(\boldsymbol{k}, \boldsymbol{q}, \omega_n\right)=0
\end{equation}
where $\delta_{\alpha \beta}$ is the Kronecker delta. We can define a second anomalous Green's function
\begin{equation}
F_{\alpha \beta}(\boldsymbol{k}, \boldsymbol{q}, \tau) \equiv-\left\langle T_\tau c_{\boldsymbol{k}+\frac{\boldsymbol{q}}{2}, \alpha}(\tau) c_{-\boldsymbol{k}+\frac{\boldsymbol{q}}{2}, \beta}(0)\right\rangle
\end{equation}
Combining the definition of the gap function, the expression of the complex conjugate of the gap function, and the definitions of the anomalous Green's function and the second anomalous Green's function in imaginary-time space and their transformations in frequency space, we can get a relationship
\begin{equation}
F_{\beta^{\prime} \alpha^{\prime}}^{\dagger}\left(\boldsymbol{k}^{\prime}, \boldsymbol{q}, \omega_n\right)=\left[F_{\alpha^{\prime} \beta^{\prime}}\left(\boldsymbol{k}^{\prime}, \boldsymbol{q}, \omega_n\right)\right]^*
\end{equation}
Then, we can replace the anomalous Green's functions by the second anomalous Green's functions in the Gor'kov equations to get
\begin{equation}
\left(-i \omega_n+\frac{\varepsilon_{\boldsymbol{k}+\frac{\boldsymbol{q}}{2}}}{\hbar}\right) G_{\alpha \beta}\left(\boldsymbol{k}, \boldsymbol{q}, \omega_n\right)-\frac{1}{\hbar} \sum_{\alpha^{\prime \prime}} \Delta_{\alpha \alpha^{\prime \prime}}(\boldsymbol{k}, \boldsymbol{q})\left[F_{\beta \alpha^{\prime \prime}}\left(\boldsymbol{k}, \boldsymbol{q}, \omega_n\right)\right]^*=-\delta_{\alpha \beta}
\end{equation}
\begin{equation}
\left(-i \omega_n-\frac{\varepsilon_{-\boldsymbol{k}+\frac{\boldsymbol{q}}{2}}}{\hbar}\right)\left[F_{\beta \alpha}\left(\boldsymbol{k}, \boldsymbol{q}, \omega_n\right)\right]^*-\frac{1}{\hbar} \sum_{\alpha^{\prime \prime}} \Delta_{\alpha^{\prime \prime} \alpha}^*(\boldsymbol{k}, \boldsymbol{q}) G_{\alpha^{\prime \prime} \beta}\left(\boldsymbol{k}, \boldsymbol{q}, \omega_n\right)=0
\end{equation}
The Gor'kov equations can be written in a matrix form
\begin{equation}
\left(-i \hbar \omega_n+\varepsilon_{\boldsymbol{k}+\frac{\boldsymbol{q}}{2}}\right) \hat{G}-\hat{\Delta} \hat{F}^{\dagger}=-\hbar \sigma_0
\end{equation}
\begin{equation}
\left(-i \hbar \omega_n-\varepsilon_{-\boldsymbol{k}+\frac{\boldsymbol{q}}{2}}\right) \hat{F}^{\dagger}-\hat{\Delta}^{\dagger} \hat{G}=0
\end{equation}
where $\hat{\Delta}$ is the gap matrix and $\sigma_0$ is the identity matrix. Solving this system of matrix equations, we obtain the Green's function and the anomalous Green's function as
\begin{equation}
\hat{G}=-\hbar\left(\varepsilon_{-\boldsymbol{k}+\frac{\boldsymbol{q}}{2}}+i \hbar \omega_n\right)\left[\left(\varepsilon_{-\boldsymbol{k}+\frac{\boldsymbol{q}}{2}}+i \hbar \omega_n\right)\left(\varepsilon_{\boldsymbol{k}+\frac{\boldsymbol{q}}{2}}-i \hbar \omega_n\right) \sigma_0+\hat{\Delta} \hat{\Delta}^{\dagger}\right]^{-1}
\end{equation}
\begin{equation}
\hat{F}=\hbar\left[\left(\varepsilon_{-\boldsymbol{k}+\frac{\boldsymbol{q}}{2}}-i \hbar \omega_n\right)\left(\varepsilon_{\boldsymbol{k}+\frac{\boldsymbol{q}}{2}}+i \hbar \omega_n\right) \sigma_0+\hat{\Delta} \hat{\Delta}^{\dagger}\right]^{-1} \hat{\Delta}
\end{equation}
For singlet pairing, the gap matrix $\hat{\Delta}(\boldsymbol{k}, \boldsymbol{q})$ is anti-symmetric which can be characterized by a single even function $\psi(\boldsymbol{k}, \boldsymbol{q})$,
\begin{equation}
\hat{\Delta}(\boldsymbol{k}, \boldsymbol{q})=\psi(\boldsymbol{k}, \boldsymbol{q}) i \sigma_2=\left[\begin{array}{cc}
0 & \psi(\boldsymbol{k}, \boldsymbol{q}) \\
-\psi(\boldsymbol{k}, \boldsymbol{q}) & 0
\end{array}\right]
\end{equation}
Substituting $\hat{\Delta} \hat{\Delta}^{\dagger}=|\psi|^2 \sigma_0$ into $\hat{G}$ and $\hat{F}$, we get the explicit forms of the Green's function and the anomalous Green's function
\begin{equation}
\hat{G}=-\frac{\hbar\left(\varepsilon_{-\boldsymbol{k}+\frac{\boldsymbol{q}}{2}}+i \hbar \omega_n\right) \sigma_0}{\left(\varepsilon_{-\boldsymbol{k}+\frac{\boldsymbol{q}}{2}}+i \hbar \omega_n\right)\left(\varepsilon_{\boldsymbol{k}+\frac{\boldsymbol{q}}{2}}-i \hbar \omega_n\right)+|\psi(\boldsymbol{k}, \boldsymbol{q})|^2}
\end{equation}
\begin{equation}
\hat{F}=\frac{\hbar \hat{\Delta}}{\left(\varepsilon_{-\boldsymbol{k}+\frac{\boldsymbol{q}}{2}}-i \hbar \omega_n\right)\left(\varepsilon_{\boldsymbol{k}+\frac{\boldsymbol{q}}{2}}+i \hbar \omega_n\right)+|\psi(\boldsymbol{k}, \boldsymbol{q})|^2}
\end{equation}
For triplet pairing, the gap matrix $\hat{\Delta}(\boldsymbol{k}, \boldsymbol{q})$ is symmetric which can be parametrized by an odd vector function $\boldsymbol{d}(\boldsymbol{k}, \boldsymbol{q})$,
\begin{equation}
\hat{\Delta}(\boldsymbol{k}, \boldsymbol{q})=\boldsymbol{d}(\boldsymbol{k}, \boldsymbol{q}) \cdot \boldsymbol{\sigma} i \sigma_2=\left[\begin{array}{cc}
-d_x(\boldsymbol{k}, \boldsymbol{q})+\boldsymbol{i} d_y(\boldsymbol{k}, \boldsymbol{q}) & d_z(\boldsymbol{k}, \boldsymbol{q}) \\
d_z(\boldsymbol{k}, \boldsymbol{q}) & d_x(\boldsymbol{k}, \boldsymbol{q})+\boldsymbol{i} d_y(\boldsymbol{k}, \boldsymbol{q})
\end{array}\right]
\end{equation}
where the Pauli matrix vector $\boldsymbol{\sigma}=\left(\sigma_1, \sigma_2, \sigma_3\right)$. Assuming real $\boldsymbol{d}(\boldsymbol{k}, \boldsymbol{q})$, we have $\hat{\Delta} \hat{\Delta}^{\dagger}=|\boldsymbol{d}|^2 \sigma_0$. Substitute this into the solutions for $\hat{G}$ and $\hat{F}$, we get the explicit forms for the Green's functions 
\begin{equation}
\hat{G}=-\frac{\hbar\left(\varepsilon_{-\boldsymbol{k}+\frac{\boldsymbol{q}}{2}}+i \hbar \omega_n\right) \sigma_0}{\left(\varepsilon_{-\boldsymbol{k}+\frac{\boldsymbol{q}}{2}}+i \hbar \omega_n\right)\left(\varepsilon_{\boldsymbol{k}+\frac{\boldsymbol{q}}{2}}-i \hbar \omega_n\right)+|\boldsymbol{d}(\boldsymbol{k}, \boldsymbol{q})|^2}
\end{equation}
\begin{equation}
\hat{F}=\frac{\hbar \hat{\Delta}}{\left(\varepsilon_{-\boldsymbol{k}+\frac{\boldsymbol{q}}{2}}-i \hbar \omega_n\right)\left(\varepsilon_{\boldsymbol{k}+\frac{\boldsymbol{q}}{2}}+i \hbar \omega_n\right)+|\boldsymbol{d}(\boldsymbol{k}, \boldsymbol{q})|^2}
\end{equation}

\end{document}